Space Project for Astrophysical and Cosmological Exploration (*SPACE*), an ESA stand-alone mission and a possible contribution to the Origins Space Telescope

Denis Burgarella[1], Andrew Bunker[2], Rychard Bouwens[3], Laurent Pagani[4], Jose Afonso[5], Hakim Atek[6], Marc Audard[7], Sylvie Cabrit[4], Karina Caputi[8], Laure Ciesla[1], Christopher Conselice[9], Asantha Cooray[10], Giovanni Cresci[11], Mirko Curti[12], Jose Miguel Rodriguez Espinosa[13], Marc Ferrari[1], Chiaki Kobayashi[14], Nadège Lagarde[15], Jesus Gallego Maestro[16], Roberto Maiolino[12], Katarzyna Malek[17], Filippo Mannucci[11], Julien Montillaud[15], Pascal Oesch[7], Chris Pearson[18], Agnieszka Pollo[17], Céline Reylé[15], David Rosario[19], Itsuki Sakon[20], Daniel Schaerer[7], Ray Sharples[21], David Sobral[22], Frédéric Zamkotsian[1]

[1] Aix Marseille Univ, CNRS, CNES, LAM, Marseille, France, ORCID: 0000-0002-4193-253, denis.burgarella@lam.fr;
[2] Department of Physics, University of Oxford, Keble Road, Oxford OX13RH, United Kingdom, andy.bunker@physics.ox.ac.uk;
[3] Leiden Observatory, Leiden University, PO Box 9513, 2300 RA Leiden, The Netherlands
[4] LERMA & UMR8112 du CNRS, Observatoire de Paris, PSL University, Sorbonne Universités, CNRS, F-75014 Paris, France;
[5] Instituto de Astrofísica e Ciências do Espaço, Faculdade de Ciências, Universidade de Lisboa, OAL, Tapada da Ajuda, PT1349-018 Lisboa, Portugal;
[6] Sorbonne Université, CNRS UMR 7095, Institut d'Astrophysique de Paris, 98 bis bvd Arago, 75014, Paris, France
[7] Observatoire de Geneve, Université de Geneve, 51 Ch. des Maillettes, 1290 Versoix, Switzerland;
[8] Kapteyn Astronomical Institute, University of Groningen, 9700 AV Groningen, The Netherlands
[9] Centre for Astronomy and Particle Theory, University of Nottingham, University Park, Nottingham, NG7 2RD, United Kingdom;
[10] University of California, Irvine, Irvine, CA, USA;
[11] INAF - Osservatorio Astrofisico di Arcetri, largo E. Fermi 5, 50127 Firenze, Italy;
[12] Kavli Institute for Cosmology - University of Cambridge, Madingley Road, Cambridge, CB3 0HA, United Kingdom
[13] Instituto de Astrofısica de Canarias (IAC), E-38205 La Laguna, Spain;
[14] School of Physics, Astronomy and Mathematics, Centre for Astrophysics Research, University of Hertfordshire, College Lane, Hatfield AL10 9AB, UK
[15] Institut UTINAM, CNRS UMR6213, Univ. Bourgogne Franche-niversidaComté, OSU THETA Franche-Comté-Bourgogne, Observatoire de Besançon, BP 1615, 25010 Besançon Cedex, France;
[16] Departamento de Física de la Tierra y Astrofísica, Facultad CC Físicas, Instituto de Física de Partículas y del Cosmos, IPARCOS, Universidad Complutense de Madrid, 28040 Madrid, Spain;
[17] National Centre for Nuclear Research, ul. Hoza 69, 00-681 Warszawa, Poland;
[18] RAL Space, STFC Rutherford Appleton Laboratory, Didcot, Oxfordshire, OX11 0QX, United Kingdom;
[19] Centre for Extragalactic Astronomy, Durham University, Department of Physics, Durham University, South Road, Durham DH1 3LE, United Kingdom;
[20] Department of Astronomy, Graduate School of Science, The University of Tokyo, Tokyo 113-0033, Japan;
[21] Center for Advanced Instrumentation, Department of Physics, Durham University, South Road, Durham DH1 3LE, United Kingdom;
[22] Department of Physics, Lancaster University, Lancaster LA1 4YB, United Kingdom

Abstract

We propose a new mission called Space Project for Astrophysical and Cosmological Exploration (*SPACE*) as part on the ESA long term planning Voyage 2050 programme. *SPACE* will study galaxy evolution at the earliest times, with the key goals of charting the formation of the heavy elements, measuring the evolution of the galaxy luminosity function, tracing the build-up of stellar mass in galaxies over cosmic time, and finding the first super-massive black holes (SMBHs) to form. The mission will exploit a unique region of the parameter space, between the narrow ultra-deep surveys with HST and JWST, and shallow wide-field surveys such as Roman Space Telescope and EUCLID, and should yield by far the largest sample of any current or planned mission of very high redshift galaxies at z > 10 which are sufficiently bright for detailed follow-up spectroscopy. Crucially, we propose a wide-field spectroscopic near-IR + mid-IR capability which will greatly enhance our understanding of the first galaxies by detecting and identifying a statistical sample of the first galaxies and the first SMBH, and to chart the metal enrichment history of galaxies in the early Universe – potentially finding signatures of the very first stars to form from metal-free primordial gas. The wide-field and wavelength range of *SPACE* will also provide us a unique opportunity to study star formation by performing a wide survey of the Milky Way in the near-IR + mid-IR.

This science project can be enabled either by a stand-alone ESA-led M mission or by an instrument for an L mission (with ESA and/or NASA, JAXA and other international space agencies) with a wide-field (sub-)millimetre capability at λ > 500 µm.

1. Introduction

There has been enormous progress over the past decade in discovering galaxies which existed early in the history of the Universe (within a billion years of the Big Bang, at z > 6), thanks in large part to images from the Hubble Space Telescope, and confirming spectroscopy from large telescopes on the ground. The next few years will see the "high redshift frontier" pushed even further with the James Webb Space Telescope (JWST) and ground based Extremely Large Telescopes (ELTs).

However, the limited field of view of these facilities (especially JWST), and sensitivity only out to the near-infrared (near-IR, λ < 2µm) for the Roman Space Telescope (formerly WFIRST) and EUCLID wide-field imaging space missions, mean that a crucial piece of the jigsaw remains missing: a wide-field imaging survey, working at near + mid-IR wavelengths (necessarily from space) is needed to find the very rare most massive and luminous galaxies at the highest redshifts, the progenitors of which are likely to be the first galactic structures to form. NIR spectroscopy at λ > 2µm (corresponding to the rest-frame optical frame) is also mandatory to get complete information (metallicity, stellar mass) on galaxies at z > 10.

The landscape of astrophysics in the timeframe from 2035-2050 is expected to be very rich: the JWST mission will have been completed, presumably finding a wealth of faint galaxies at high redshift and addressing the role of these early galaxies in the reionization of the inter-galactic medium. ALMA will be a very mature facility by then and SKA will have explored the molecular emission and dust re-emission from some of these objects. The re-ionisation of the Universe was achieved by low luminosity sources (Bunker et al. 2004, 2010, Ouchi et al. 2010, Bouwens et al. 2010, Robertson et al. 2015). These low luminosity sources would only be visible if they are in groups or proto clusters (Castellano et al. 2016). This is likely so for the first galaxies, which were of very low luminosity. Thus, detecting proto clusters from z~6 to z~15 would unveil the history of the Universe's re-ionisation (Calvi et al. 2019; Rodriguez Espinosa et al. 2019). Rare and bright sources at high redshift (as well as transients such as distant supernovae) will be explored by LSST on the ground, and EUCLID and the Roman Space Telescope in space, at wavelengths below 2 microns. In the X-ray, after a hiatus of many decades new facilities such as ATHENA will see AGN out to unprecedented distances. But there is a key gap in the parameter space that remains unexploited - a wide-field IR survey mission with spectroscopy and imaging working beyond 2 microns that we propose to address here.

Our proposed *SPACE* mission will place the ESA community in a leading position to study the early universe after JWST's deep pencil beam surveys via an unbiased census of the first-light objects. The SPACE mission will also provide us with the high angular resolution survey of the Milky Way in the near-IR + mid-IR to statistically study the star formation phenomenon in the Milky Way. *SPACE*'s wide-field imaging and integral-field / multi-object spectroscopy in space are the major instrumental breakthroughs that will enable this new window in astronomy.

2. Extragalactic Astrophysics and Cosmology

Our main objectives for extragalactic astrophysics are related to the birth of the first galaxies and AGN, and to the rise of the metals in the Universe.

2.1. Key questions

2.1.1. When and How Did Galaxies Form?

The *SPACE* mission will achieve photometric identification of primordial star–forming galaxies at 10 < z < 15, i.e. less than 0.3 - 0.5 Gyr after the Big Bang, over much larger volumes than available to HST or JWST. The galaxies will be selected through the Lyman break technique (e.g. Steidel et al. 1996, Bunker et al. 2004, Burgarella et al. 2007, Bouwens et al. 2010) where intervening neutral hydrogen absorbs all light at wavelengths short-ward of Lyα1216Å in the rest-frame, causing the highest-redshift objects to "drop out" of visible and near-IR images. The expected density of these galaxies at z > 14 is estimated to be ~1 deg$^{-2}$ at m$_{AB}$ = 28 (e.g. Burgarella et al., https://arxiv.org/pdf/1607.06606.pdf). To detect a statistically significant sample at the highest redshifts, an imaging survey of about 200 deg$^2$ is necessary. Several hundred sources are needed so as to determine the luminosity function and its evolution with redshift (particularly the shape at the bright end to probe the effects of feedback at early times). This sample of luminous galaxies will be the largest one obtained with any facility because: (1) JWST is unlikely to build surveys over areas much larger than ~1 deg$^2$, i.e., HST-like which means that the probability of detecting such luminous galaxies is low at z > 12 and, (2) detecting these primordial galaxies requires any facility to have at least 2 bands at λ > 1.3µm (at z = 10) and λ > 2.0µm (at z = 15), to provide robust colours at wavelengths beyond the Lyman break for photometric redshifts. To date, only JWST and *SPACE* have the appropriate wavelength range extending beyond 2.0µm with the required sensitivity.

### 2.1.2. Spectroscopic detection and identification of photometrically faint emission line galaxies via a blind spectroscopic survey (with integral-field spectroscopy)

Imaging surveys allow the detection of galaxies with a strong continuum. However, an important subset of galaxies, younger (and therefore at low mass compared to today's standard) and undergoing strong star-bursting events, are optimally detected via spectroscopic surveys aiming at strong emission lines. VLT MUSE's results (Bacon et al. 2015) discovered many Lyα emitting galaxies that are completely undetected in the HST deep images down to I(F814W) > 29.5. These sources that have no HST counterparts represent 30% of the entire Lyα emitter sample. A blind and relatively wide-field integral-field spectroscopic (IFS) survey is the unique way to detect these objects – and provides a great sensitivity advantage over "slitless" spectroscopy, even from space, where each pixel records background noise at all wavelengths. No other current or planned space facility will feature an IFS instrument with such a large field-of-view. *SPACE*'s IFS will build a survey via parallel observations and reach magnitudes as deep as the widest JWST NIRSpec survey (e.g. in the JWST Wide Field survey as defined in Mason, Trenti & Treu 2015, in the F115W and F220W filters) but over a much larger area of more than 1deg$^2$ (ten times the area over which the anticipated NIRSpec surveys will draw their targets from). Moreover, JWST/NIRSpec will not obtain spectroscopy over its entire field of view (except in the NIRCAM and NIRISS slitless grism mode but still on small fields of about 2.2' x 2.2'): a prior photometric detection is needed to define the slits for the micro-shutter arrays (and the alternative IFS mode on NIRSpec covers 3x3arcsec$^2$, only 0.03% of the full field of the NIRSpec micro-shutter array). IFS with *SPACE* will look for signatures of the first generation of Population III stars in the earliest galaxies, chart the evolution of metal enrichment and the assembly of mass in galaxies over cosmic time.

### 2.1.3. The first quasars and massive black holes

The density of very high redshift (z > 6) luminous quasars is very low, with only a handful currently known at z > 7, although the very existence of supermassive black holes at these early times presents strong challenges to seed formation and black-hole growth models. Crucially, *SPACE* has the wavelength coverage to identify quasars out to high redshift, and *SPACE*'s field-of-view is large enough to directly detect ~100 quasars at z > 8 pushing down to fainter luminosities than current surveys. Moreover, we will provide sufficient time in *SPACE*'s observation schedule to observe photometrically and spectroscopically these quasars into the rest-frame optical. This will provide unique information on the early co-evolution of galaxies and super-massive black holes but also, this will allow the study of the intergalactic medium along the line of sight. The IFS will be very valuable to try and detect the environment of these early super-massive black holes.

### 2.1.4. The birth of metals

Big bang nucleosynthesis creates hydrogen and helium, but all elements heavier than beryllium are formed later in stars and supernovae. Hence the "metallicity" (the enrichment of cosmic gas by heavy elements) acts as a clock - the elemental abundances are built up with time with successive generations of star formation. Emission line diagnostics have been used at intermediate redshifts to determine the metallicity of the star- forming gas, and a key probe of galaxy evolution is the mass-metallicity relation (e.g. Tinsley 1980), which indicates how star formation and chemical enrichment proceed in galaxies as a function of galaxy stellar mass. Good determinations have been made at low redshift (e.g. SDSS, Tremonti et al. 2004, Mannucci et al. 2010) and more recent work at intermediate redshifts (e.g. Zahid et al. 2011 at z~1). Metallicity measurements are currently limited to z ≲ 3 due to spectral coverage from the ground, and there exists a small number of z ~ 3 estimates (e.g. Maiolino et al. 2008, Maiolino & Mannucci 2019). The slope and offset of the mass-metallicity relation will evolve with redshift if the characteristic timescale (or efficiency) of chemical enrichment depends on the stellar mass. At the higher redshifts there appears to be significant differences between the current small number of observations and theoretical predictions from chemical evolution models (e.g. Taylor & Kobayashi 2016, Davé et al. 2011).

## 2.2. Detailed science case

### 2.2.1. The First Galaxies

Our understanding of how galaxies form and evolve over the last 13 Gyrs of cosmic time has increased dramatically over the last decades (e.g. see reviews Robertson et al. 2010, Dunlop 2012; Madau & Dickinson 2014). Tremendous effort has been invested into finding and studying the most distant sources, particularly since they offer stringent constraints on galaxy formation, allowing us to test models of early galaxy formation and evolution (e.g. Vogelsberger et al. 2014; Schaye et al. 2015; Lacey et al. 2016), along with studying the epoch of reionization (e.g. Shapiro et al. 1994; Iliev et al. 2006; Dijkstra 2014). Nevertheless, some fundamental questions remain regarding the very early Universe. Where and when did the first stars and black holes form? What were their properties? Can we finally identify and characterise them, thus confronting state-of-the-art predictions and pushing models much further? While ultra-deep surveys with HST have been extremely successful, most of the small number of current z = 10 candidates have not been found in the Hubble Ultra Deep Field but in wider-field, shallower HST surveys. We now have growing evidence that the bright end of the luminosity function is evolving less rapidly than the faint end (e.g. Bowler et al. 2014), which is not well captured or reproduced in current models. This motivates a search for bright galaxies out to high redshift to accurately determine this evolution. The unique combination of depth, area, and wavelength coverage provided by the imaging camera on *SPACE* will provide the first significant searches for z > 10 galaxies (potentially revealing hundreds of high redshift sources with star formation rates greater than 100 $M_\odot yr^{-1}$), for the first quasars at z > 7, and for massive evolved galaxies at z > 6. With these searches, we can constrain unexpected physics and test the $\Lambda$CDM paradigm. Complementary to this imaging, the wide-field integral field spectroscopy (IFS) capability of *SPACE* offers an unprecedented opportunity to study very young and low-mass strongly star-bursting proto-galaxies in the early Universe, to search for the elusive first generation of stars (Population III), and to chart the chemical enrichment history and mass assembly of galaxies at early times.

**The high redshift frontier**: The brightest most massive galaxies at high redshift occupy a unique region in parameter space which provides a new test of galaxy formation physics in a different regime to that probed by the small-area surveys of HST and JWST. At z = 13-15, the universe will have an age of only 250 - 300 million years. These early times are as yet unexplored, and it is unlikely that JWST will survey sufficient area to identify even one such galaxy near the bright end of the luminosity function (LF), while surveys like Euclid or WFIRST are limited to λ < 2µm and do not have sufficiently red filters to identify such galaxies at all (Fig. 1). Pushing to these hitherto-unexplored highest redshifts is key to understanding when the first stars and galaxies formed.

**Large samples of robust candidates at z ~ 11-12**: *SPACE* is also likely to provide the community with the largest and most complete sample of ultra-bright galaxies at z = 11-12. While sources at these redshifts can, in principle, be selected based on the WFIRST survey (the EULID survey will likely be too shallow at z > 10), rare z~11-12 galaxies will be difficult to distinguish from the much more ubiquitous contaminant population due to the lack of filters beyond 2µm (and hence restricted colour information), making any confirmation campaign extremely time expensive. The *SPACE* mission overcomes these significant issues with contamination and expensive follow-up confirmation.

Most of the high redshift objects lacks spectroscopic redshifts, with the luminosity functions derived just from LBG candidates identified in images. Accurate spectroscopic redshifts are critical, since the interloper fraction of lower-redshift sources is very uncertain. Critically, the inferred luminosity function depends strongly on the calculation of the volume sampled, which in turn depends on the selection and completeness (Fig. 2). These are strongly affected by the redshift distribution and spectral slopes of the Lyman-break galaxies.

In order to effectively select Lyman break galaxies at high redshift, we will use optical imaging to reject low redshift red contaminants in our fields (which would be detected at shorter wavelength, unlike the highest redshift objects). The planned LSST deep-drilling fields will cover ~50-100 $deg^2$ and could be expanded further to match our intended survey area. The anticipated depths of $m_{AB}$ = 28 at 5σ (equivalent to a 2σ upper limit of $m_{AB}$ = 29) are sufficient to cleanly select Lyman break galaxies with $m_{AB}$ = 28 at wavelengths beyond Lyman-alpha, particularly when the *ugrizy* images are co-added for greater depth. Alternatively, the existing HyperSuprimeCam on Subaru will reach comparable depths in *g* and *r* band in the planned Ultra Deep survey (2.4 nights per pointing). A survey in a single band with Subaru over *SPACE*'s 200$deg^2$ area would take about 260 nights.

Making use of published predictions from Mason et al. (2015), Behroozi & Silk (2015) and Mashian et al. (2016), we predict we would find ~5000 z ~ 13 galaxies, ~200 at z ~ 14 galaxies, and still a few tens galaxies at z ~ 15 (in redshift slices of Δz = 1) down to our limiting magnitude of $m_{AB}$ = 28, compared with at most a few tens of galaxies at this brightness for JWST (Fig. 3) and virtually none such sources from the EUCLID and WFIRST missions. Several tens thousands z ~ 11 and z ~ 12 ultra-bright galaxies are also predicted for the *SPACE* mission, dwarfing what are likely to be limited samples with WFIRST+EUCLID + potential follow-up campaigns with JWST (Fig. 3). These numbers are uncertain given the current weak constraints on the luminosity function at high redshift, hence the need for a large *SPACE* sample to accurately determine the evolution of the LF. If the bright end evolves less rapidly than in these models, these predicted numbers will be lower limits.

The large samples of z = 11-15 galaxies from the *SPACE* mission will be enormously valuable, allowing for a wide variety of follow-up endeavors (spectroscopy from ELTs, X-ray observations with Athena and radio/mm with SKA and ALMA). A recent work detected a candidate for the remote object in the universe, at z = 11.1 (Oesch et al. 2016). This galaxy is remarkably, and unexpectedly, luminous for a galaxy at such an early time but also very rare. If confirmed, this result implies that the best strategy to detect z > 10 is wide fields, as featured by *SPACE*. With these samples, *SPACE* will provide the community with new constraints on the prevalence of ultra-bright z ~ 13-15 galaxies and more robust constraints at z ~ 11-12. This can be used to assess the general form of the UV luminosity function at z > 10. There is tentative evidence based on recent observations for the galaxy LF becoming a double power-law shape at z > 6, compared with a Schechter function with an exponential cut-off at redshifts lower than this (Bowler et al. 2014). One possible explanation for this is that the most UV luminous galaxies probed at the highest redshifts actually have lower stellar masses than the same UV luminosity galaxies at lower redshifts, due to both dust extinction (Bogdanoska & Burgarella 2020) and AGN feedback being lower at the high redshifts. This might result in the bright end of the UV LF at higher redshifts increasingly resembling the mass function (a power law) rather than a Schechter function.

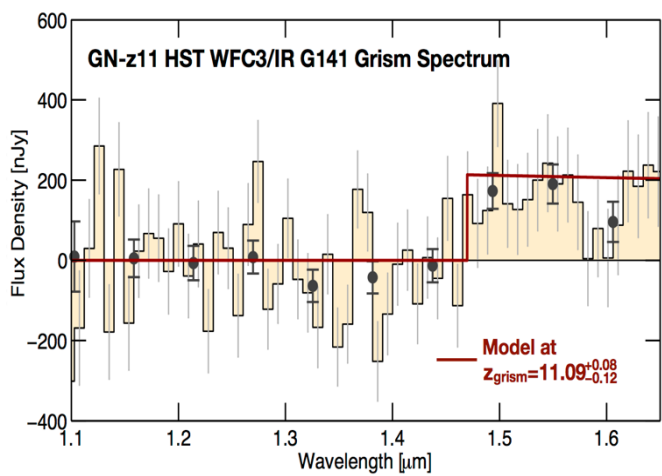

Figure 1: The Spectral Energy Distribution of the galaxy discovered at z = 11.09 (Oesch et al. 2016) presents a break around λ = 1.5μm due to the intervening Lyman-alpha forest absorption. To get a reliable estimate of the redshifts in the 10 < z < 15 range, it is crucial to collect data at wavelengths above and below the break. The NIR+MIR range is therefore needed. This is JWST's choice and also that of *SPACE*. Other facilities with limited capabilities beyond ~2μm will not be suitable for this quest – two or more filters longward of the break are needed for secure identification. Credit: P. Oesch / Observatoire de Genève.

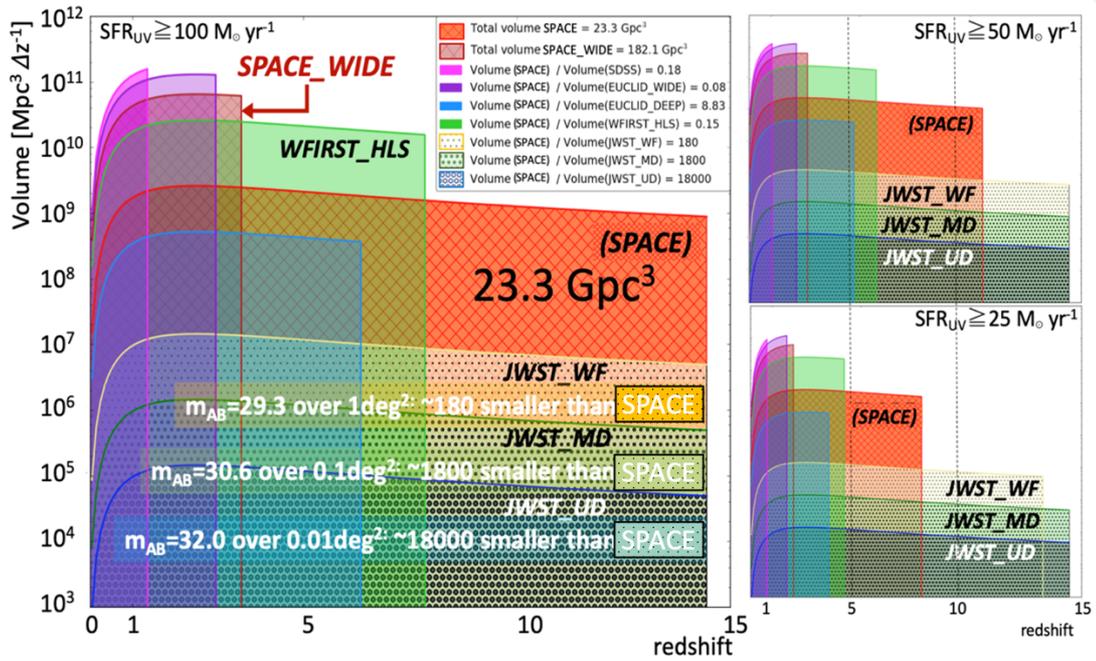

Figure 2: The volume of the universe for different projects in which galaxies with various star formation rate thresholds (from the rest-frame UV) can be detected. The legend gives *SPACE*'s volume and the ratio of the other projects with respect to *SPACE*. None of the JWST surveys will have comoving volumes comparable to lower redshift ones (z < 6). On the other hand, *SPACE*'s 200deg$^2$ survey at z > 8 will reach about 10% the volume of the EUCLID WIDE and WFIRST HLS surveys at much lower redshift. Even at lower SFRs, i.e. normal star-forming galaxies, *SPACE* will be able to cover a huge volume unreachable for any other survey.

While such issues will almost certainly be resolved at z < 10 by JWST, Euclid, and WFIRST, galaxy formation at z = 11-15 faces several unique challenges, the largest of which is the short time scale available for the build-up of the stellar population in a galaxy at such high redshifts. At these redshifts, there is only ~100 million years (Myr) available from the formation of the first star to that galaxy forming tens of millions of stars per Myr. This is a very short time scale, particularly in relation to the time scale on which various physical processes operate, i.e., metal enrichment, feedback from supernovae, etc., and it is possible that galaxy formation proceeds differently in this regime. *SPACE* observations at these very high redshifts are crucial to constrain this.

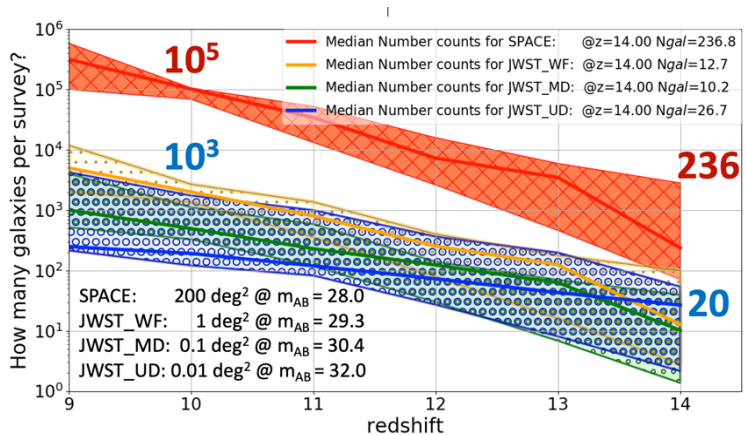

Figure 3: Number of galaxies to z = 14, detected in *SPACE* 200deg$^2$ and the three JWST surveys over 1deg$^2$, 0.1 deg$^2$ and 0.01 deg$^2$ (as defined in Mason et al. 2015). The combined depth / area gives about the same number of objects for each of the JWST surveys. With the same assumptions, *SPACE* will provide about 240 objects while JWST surveys would have about 20 each. This is a huge gain to understand these galaxies in a statistical way.

The unique samples of ultra-bright z = 11-15 sources discovered by *SPACE* will have utility that goes well beyond just studies of their prevalence, as they are extremely high value follow-up targets. Targeting very high redshift galaxies from *SPACE* with NIR imagers on the next generation of Extremely Large Telescopes will give us a direct view of star formation in the most significant mass reservoirs in the early universe, measuring their size and structural properties. Ultra-bright galaxies out to z ~ 7 have been shown to have a complex, multi-component morphology suggesting merging activity, and it would be extremely interesting to know if such phenomena would also be present in ultra-bright z ~ 11-15 galaxies building up in ≲100 Myr.

There are fascinating insights to be gained about the dark-matter – galaxy connection at the bright end through studying clustering. These bright galaxies are likely the sites of the highest over density peaks in the early Universe, and thus expected to be clustered. *SPACE* may find proto clusters, the earliest large-scale structures to form.

Follow-up spectroscopy of very high redshift galaxies only discoverable by *SPACE* (including with the *SPACE* integral field spectrograph (a multi-object spectrograph can also be considered) and from NIR spectrographs on ELTs) will enable us to determine the physical state of ionized gas (the gas-phase metallicity, electron density and ionizing flux). This is critical, since very little is known about the physical conditions of star formation in the early universe. While huge progress will be made in probing these conditions in lower luminosity galaxies with JWST, JWST will likely not look at many ultra-bright galaxies at z > 12 with $SFR_{UV}$ > a few tens $M_\odot yr^{-1}$ due to the challenge in finding them prior to *SPACE*.

2.2.2. The birth of metals

A key probe of galaxy evolution is the mass-metallicity relation (e.g. Tinsley 1980), which indicates how star formation and chemical enrichment proceed in galaxies as a function of mass. If the characteristic timescale (or efficiency) of chemical enrichment depends on the stellar mass, then one would expect the shape (slope and offset) of the mass-metallicity relation to change with redshift. Good determinations have been made at low redshift thanks to the SDSS (e.g. Tremonti et al. 2004) and more recent work at intermediate redshifts (e.g. Zahid et al. 2011 at z ~ 1 from DEEP2 on Keck). A small number of galaxies at z ~ 3 also have metallicity estimates (e.g. AMAZE on VLT, Maiolino et al. 2008). At the higher redshifts there appear to be significant differences between the current small number of observations and theoretical predictions from chemical evolution models (e.g. Taylor & Kobayashi 2016, Davé et al. 2011). It is important to improve the observations (in both number and the mass range covered) and to extend the samples to higher redshifts. *SPACE* will achieve these observations and be critical in constraining galaxy evolution models (in particular feedback).

The fine structure lines in the mid-IR and far-IR are very powerful to measure the metallicity of massive and high-metallicity galaxies (e.g., Fernandez-Ontiveros et al. 2016). They do not suffer from the dust extinction and from the degeneracy observed for high-metallicity galaxies when using rest-frame optical lines. However, galaxies are expected to have lower masses and low metallicities at z > 5 (e.g. Torrey et al. 2018, Burgarella et al. 2020, Nanni et al. 2020). If we want to focus on the rise of metals in the universe at z > 5, these IR fine structure lines are not good tracers for low-metallicity galaxies. This is a direct implication of the galaxy formation process in the $\Lambda$CDM cosmology (Fig. 4):

- Galaxies build stellar mass with time
- Higher redshift galaxies have lower mass (M < $10^{10.5}$ $M_\odot$)
- We know there is a mass – metallicity relation
- Lower mass galaxies have lower metallicities (Z < 0.2 $Z_\odot$ or 12+log(O/H) < 8.4)
- Higher redshift galaxies have lower metallicities

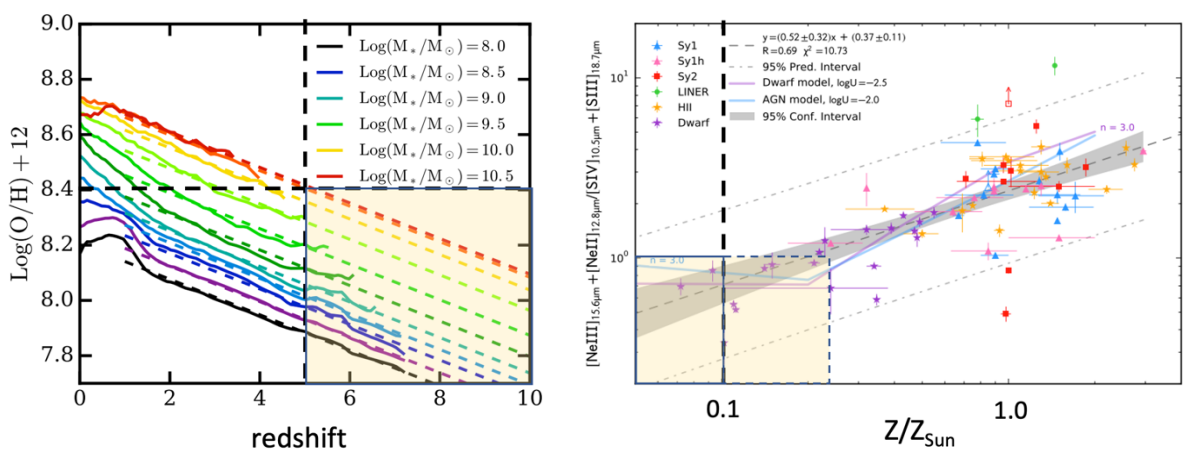

Figure 4: Even though exceptions exist, the left panels (adapted from Torrey at al. 2018) shows that most galaxies are z > 5 are expected to be low-mass galaxies at low metallicity (yellow-shaded area). In this metallicity range, the IR fine structure line reach a plateau and are not sensitive to any change in metallicity as shown in the right panel adapted from Fernandez-Ontiveros et al. (2016). Credit: Left: P. Torrey / University of Florida and Right: J.M. Fernandez-Ontiveros / Instituto de Astrofisica de Canarias.

If we cannot use the fine structure line, what are the options? We can utilize the strong power of the bright rest-frame optical lines and (tentatively) make use of the 3.3μm PAH feature:

a. There is a large diversity of metallicity indicators (see Maiolino & Mannucci 2018 and Fig. 5). But, traditionally, emission line diagnostics have been used at intermediate redshifts to determine the metallicity of the star-forming gas. For many years a widely-used indicator for the O/H abundance has been "R23" (Pagel et al. 1979), which uses [OII], [OIII] and Hβ as lines which are reasonably close in wavelength (and hence minimizes the effect of differential dust extinction) and accounts to first order for ionization by using two species of oxygen (see also the updated O/H method of Curti et al. 2017). We can determine R23 out to z ~ 9 with *SPACE*. However, there is a well-known "double fork" in the plot of R23 against metallicity, but the [NII]/Hα line ratio can be used to break this degeneracy and determine if a galaxy lies on the upper or lower branch (Fig. 5). With *SPACE*, we can track [NII]6583Å out to z ~ 7, much further than the current z ~ 2.5 limit. Our *SPACE* spectroscopy will access several emission lines, enabling us to use "BPT" diagrams (Baldwin, Philips & Terlevich 1981) to address the nature of the photoionization in individual galaxies (i.e. star formation vs. AGN). The line ratios will also probe the ionization parameter, and recent work has suggested that this rises with redshift (Kewley et al. 2015 at z ~ 0.6).

b. We will also be sensitive to the poly-aromatic hydrocarbon (PAH) features in galaxies at high redshift. The PAH are destroyed in low-Z environment by the UV radiation field, which propagates more easily due to the lower dust content for galaxies with lower metallicities (Galliano et al. 2003; Boselli, Lequeux & Gavazzi, 2004; Gordon et al. 2008). The PAH3.3μm could be an excellent vector to measure the metallicity over a very wide range of redshifts, using the same indicator. The low thermal background of *SPACE* permits much deeper observation in the MIR at λ > 13μm than JWST-MIRI. In this range, we observe a strong and linear decrease of $q_{PAH}$, the mass fraction of PAH (Fig. 6) that seems to be correlated to the metallicity of the observed galaxies. Several interpretations for this effect exist. But the leading one is that PAH are destroyed in low-Z environment by the UV radiation field, which propagates more easily due to the lower dust content for galaxies with lower metallicities (Galliano et al. Boselli et al. 2004, Madden et al. 2006; Gordon et al. 2008). The PAH3.3μm will be an excellent vector to measure the metallicity over a very wide range of redshifts, using the same metallicity indicator.

Of course, there is not just one "number" which describes the chemical enrichment of an individual galaxy. We can improve on R23 metallicity by measuring abundances of different elements. This would provide strong constraints on galaxy evolution, and in particular the relative contribution to chemical enrichment from core-collapse supernovae (SNe) and asymptotic giant branch (AGB) stars (e.g. Nomoto et al. 2013), which have different timescales and hence the abundance patterns should evolve strongly with redshift. The N/O ratio can be determined using the [SII], [NII], Hβ and [OIII] lines to constrain the ionization parameter (Dopita et al. 2016), and similarly the CIII]1909 and CIV lines provide the carbon abundance (Amorin et al. 2017). Among AGB stars, nitrogen is produced from relatively massive stars (~5$M_\odot$), while carbon is produced by the low-mass end (1 - 3 $M_\odot$) with longer lifetimes. Oxygen is produced very promptly from core-collapse SNae (with >8$M_\odot$ progenitors). The relative abundances probed by *SPACE* spectroscopy tell us about the galactic archeology, and potentially any evolution of the stellar initial mass function. For the brighter sources at the medium spectral resolution (R = 600-1000) we may be able to detect absorption lines – this may constrain the stellar metallicity (rather than gas phase) and be less affected by the ionization parameter and dust. However, many of the absorption lines arise from the interstellar medium (ISM) rather than from stars. This gives information on the velocity of outflows (the feedback from star formation).

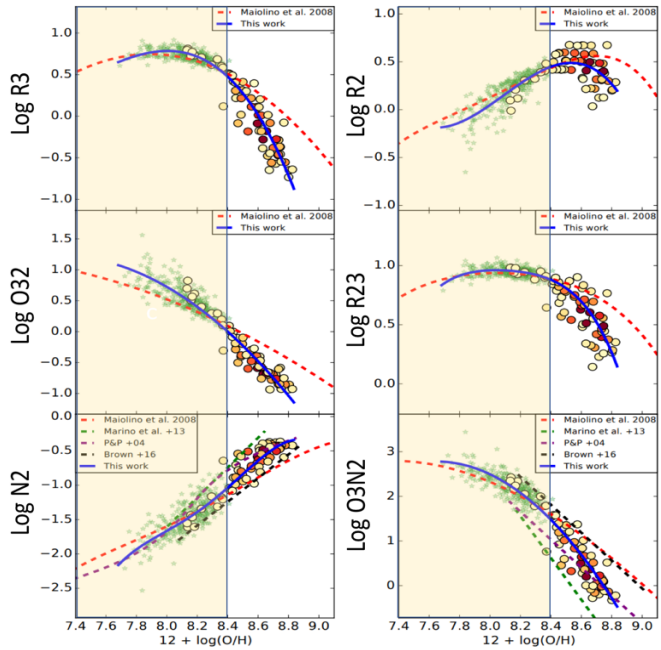

Figure 5: This plot extracted from Curti et al. (2017) shows the diagnosis using strong optical lines as a function of the oxygen abundance. Using these lines, a mid-IR spectrograph would be able to measure the metallicity of galaxies with the same tracer from z = 0 to z = 10. Some of the tracers show degeneracies while other, involving H$\alpha$ do not. The preferred tracers for the low-mass, low-metallicity galaxies (yellow region) would be O32 ([OIII] λ5007/[OII] λ3727), N2 ([NII] λ6584/H$\alpha$) and O3N2 ([OIII] λ5007/H$\beta$)/([NII] λ6584/H$\alpha$). This means that we need to be able to observe enough galaxies (mapping efficiency) in the good wavelength range, accounting for the redshift. Credit: M. Curti, G. Cresci & F. Mannucci, INAF.

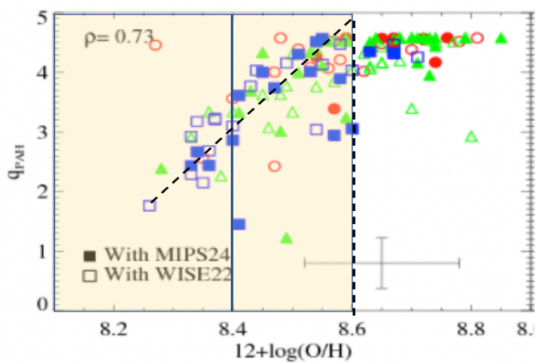

Figure 6: this plot, extracted from Ciesla et al. (2014) suggests that PAH are destroyed by the UV radiation field in the specific metallic environment of low-metallicity galaxies in the local universe and probably also in the early universe. Following up the PAH3.3 μm feature with *SPACE* will allow to better understand this behavior and to apply it to very high-redshift galaxies to measure the metallicity of these objects, with the goal of charting the rise of metals in the early times. Credit: L. Ciesla / LAM.

**Metal enrichment and supernovae at high redshift:** Probing the metallicity in ultra-bright galaxies will be key to understanding their evolution, as it could provide us with earliest high-quality information about the chemical enrichment occurring in the most massive reservoirs of star formation in the early Universe. Such probes of the metallicity are likely to be very interesting as they could provide us with information on the nature of some of the first SNe that occur in the universe and enrich the interstellar medium. This subject has huge potential, as it is thought that many of the first stars, being metal free and massive (with masses of the order of >40M$_\odot$, given the limited cooling of molecular hydrogen e.g., Bromm & Larson 2004), undergo pair-instability supernovae (PISNe). In this regard, ultra-bright z ~ 14-15 galaxies - due to both their brightness and very young ages (< 100 Myr) - arguably provide us some of the best targets to look for evidence of enrichment from such SNe in the early universe.

By observing fields repeatedly, *SPACE* may also be able to directly detect these SNe. PISNe produce huge amounts of iron (from Nickel-56) with a long plateau, and thus can be detectable even at high redshifts. The cosmic rate of PISNe depends on the formation rate of stars in the mass range 40-300M$_\odot$. In other words, the detection or non-detection of PISNe can constrain theoretical models. Whalen et al. (2013) estimate magnitudes for PISNe at z = 20-30, which can rise as bright as m$_{AB}$ = 26 mag at 2μm and m$_{AB}$ = 27 mag at 4.4μm. These are within reach of the proposed *SPACE* survey to m$_{AB}$ = 28, and hence *SPACE* has both the sensitivity and the large area mapping capability to search for PISNe. JWST will lack the field of view to make detecting this population a realistic proposition, and LSST will be limited in sensitivity at 2μm and beyond, limiting its redshift range. The light curve decline will be very long due to the time dilation at these redshifts, and we will adapt our survey strategy to re-visit target fields with the appropriate cadence (spanning months and years).

2.2.3. The Assembly of Stellar Mass in Galaxies

Parallel to the quest for the highest redshift star-forming galaxies, the *SPACE* imaging survey will also detect the redshifted starlight from the most massive galaxies over a wide range of redshifts (4 < z < 10), where the wavelength coverage of *SPACE* extends to the rest-frame optical. The presence of massive galaxies, i.e. $M_{star}$ > 5 x $10^{10}$ $M_\odot$, at early cosmic times constitutes a fundamental constraint for galaxy formation models. According to these models, no such massive galaxies are expected to have formed at z > 6, when the Universe was less than a billion years old and should become common only much later in cosmic time. This is a consequence of our current understanding of how galaxy formation proceeds in the ΛCDM cosmological model. However, different observational results are starting to challenge these predictions with the discovery of a few rare massive galaxy candidates at high redshift (e.g. Duncan et al. 2014; Caputi et al. 2015; Steinhardt et al. 2016). There is some evidence that the stellar mass – halo mass relation evolves, such that star formation is more efficient at high redshift (Behroozi 2013, Finkelstein et al. 2015, Harikane et al. 2016). But the samples are very small, and clearly much larger samples are required. This will provide crucial insights into the feedback mechanisms which regulate the assembly of stellar mass in dark matter haloes.

Given the rarity of these objects, deriving a useful constraint for galaxy formation models requires the analysis of deep galaxy surveys over wide areas of the sky covering at least several tens of square degrees. No current or forthcoming telescope can provide deep IR images over such large areas. *SPACE* will be the only telescope able to quantify the presence of massive galaxies at early cosmic times, resulting in crucial constraints for galaxy formation theories. *SPACE* will provide excellent stellar masses for huge numbers of objects in the fields, over much larger areas than JWST will be able to address. Indeed, by extrapolating the results of current galaxy surveys over small areas, we expect *SPACE* to find around 30,000 massive galaxies (> $10^{10}$ $M_\odot$) in a ~200 $deg^2$ survey at z = 6 (in a redshift slice of Δz=1), and potentially 5,000 at z = 7, along with more than 100,000 massive galaxies at z ~ 4-5. These unprecedented statistics will allow us, for the first time, to study galaxy clustering in the early Universe with the same level of detail as we currently do at z < 1 (Zehavi et al. 2011) and recently started doing at z ~3 (Durkalec et al. 2015). *SPACE* will open up a new era of 'precision cosmology' at high z, only limited by the accuracy of photometric redshifts (which will be self-calibrated with the thousands of spectroscopic redshifts that *SPACE* itself will provide). There are no current constraints on the number of massive galaxies at higher redshift, but if we assume the same shape of the galaxy stellar mass function as at lower redshift, we would expect *SPACE* to find 1,500 massive galaxies at z~8 (> $10^{10}$ $M_\odot$ and Δz = 1) and 50-100 at z ~ 9-10. The distinctive wavelength coverage of *SPACE* (1 - 28μm), which goes beyond that of any other planned wide-field facility, is uniquely able to determine the stellar masses of all these high-z galaxies (Fig. 7). The information on $M_{star}$ will provide a unique opportunity to study the growth of baryonic structures from the early Universe to the present day.

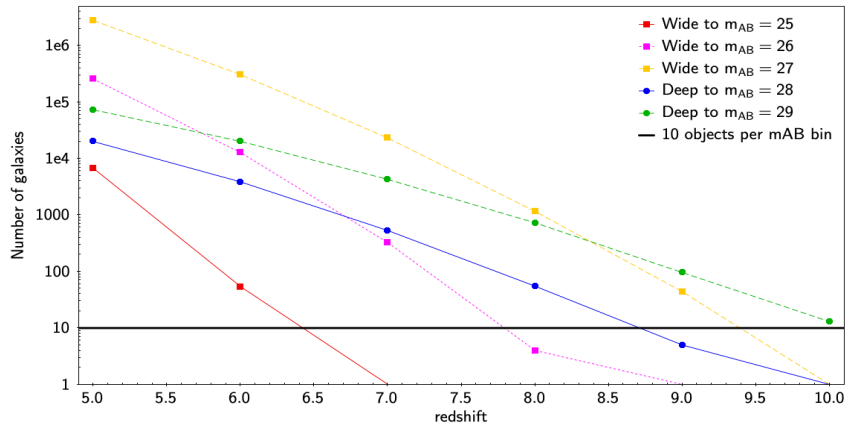

Figure 7: Number of galaxies that will be detected by the two Roman Space Telescope (formerly WFIRST) Deep and Wide surveys at 5 < z < 10. *SPACE* will be able to estimate $M_\star$ for a representative sample of at least 10 galaxies in the bins where WFIRST detected objects to z = 10. The Roman Space Telescope will provide the SFR from the rest-frame FUV.

More specifically, *SPACE* will provide a unique way of following-up WFIRST-Deep and WFIRST-Wide surveys. After WFIRST, no projects will be able to measure the star formation rate (SFR), the stellar mass ($M_\star$), and the dust attenuation of galaxies detected from WFIRST surveys until we have *SPACE* photometric capabilities. The rest-frame UV spectrum will provide an access to young stars that are likely to be predominant at z > 5 but, if we want to perform a complete census, including potential older stars, we need the rest-frame optical and near-IR. To follow up WFIRST's objects at z > 5, after JWST, we need a wide-field NIR+MIR instrument From these unique data, we will be able to measure the evolution of the mass function (cosmic mass assembly), the SFR density, average dust attenuation for a representative sample of galaxies at 5 < z <10. In 2035+, a number of facilities will have surveyed the sky to identify galaxies before and slightly after the end of reionization. JWST will certainly open up this era but galaxies are expected to be very rare and faint at 5 < z < 10. WFIRST's Deep and Wide surveys will provide a unique opportunity to build statistical samples (Tab. 1).

### 2.2.4. Formation and Growth of the First Super-Massive Black Holes

All massive galaxies today appear to harbor a central super-massive black hole. A key question is how the first massive black holes form and grow? This can be addressed by searching for Active Galactic Nuclei (AGN) at high redshift. The most massive black holes at z ~ 6-7 are ~$10^9$-$10^{10}$ M$_\odot$ which already places stringent constraints on their formation mechanism and their rate of growth from the first black-hole seeds to just z ~ 6-7. This will be achieved from *SPACE* by providing measurements of the quasar luminosity function out to z ~ 10-12 for luminosities ~5 magnitudes below the absolute magniture M* through photometric identification with broad band colours.

Spectroscopic follow-up of these candidate quasars (potentially with the *SPACE* integral field or multi-object spectrograph) will measure the black-hole accretion properties, such as the Eddington ratio (using virial estimators from broad emission-line components), as well as the metal enrichment from the strength of metal lines. A puzzling fact about the highest-redshift quasars currently known is that they are all universally metal rich, suggesting that they have already undergone a substantial amount of evolution. Spectroscopy of quasars from *SPACE* at even higher redshifts and fainter magnitudes will address whether this persists out to yet higher redshifts and for lower-luminosity systems. A drop in metal richness will have implications for the fraction of obscured systems since the obscuration is dominated by metals and we may therefore expect an increasing population of dust-free and unobscured quasars at very high redshift. *SPACE* imaging of fields surveyed by Athena (for X-ray sources) and SKA (for radio sources) will determine fraction of obscured/unobscured systems. These properties may be connected to the host-galaxy environment and the large-scale environment within which quasars reside, and the large-scale quasar environment can be measured using the large *SPACE* IFU through the identification of associated galaxies and AGN with the quasar. The predictions shown in Fig. 8 suggest that *SPACE* will detect in a 200 deg$^2$ survey a few hundred quasars at z > 6, up to ~100 at z > 8, and ~5 - 10 at z > 10. We note that there are currently only a handful of QSOs known at z > 7 (e.g., Mortlock et al. 2011, Banados et al. 2018).

We also note that in a 200 deg$^2$ survey with Athena, the predicted numbers of z > 8 and z > 10 AGN with an X-ray luminosity L$_X$ > $10^{43}$ erg/s (broadly equivalent to the rest-frame 1450Å depth of *SPACE*), are similar to those of *SPACE*. Although we would expect significant overlap in the identified source populations between Athena and *SPACE*, they will also be very complementary, with Athena being more effective at detecting the obscured systems but unable to measure redshifts (and hence source properties) for the majority, which will require *SPACE*. The timescale of Athena is well suited to follow-up *SPACE* detections, and, as well as AGN Athena could potentially target clusters of galaxies at z > 2 discovered by *SPACE* to measure the temperature of the intra-cluster medium (and hence the total cluster mass). Note that *SPACE* will also feature a wide survey to study the Milky Way. Although this survey will suffer from high MW extinctions since it is, by design, close to the disk of our galaxy, it might provide an additional sample of bright quasars.

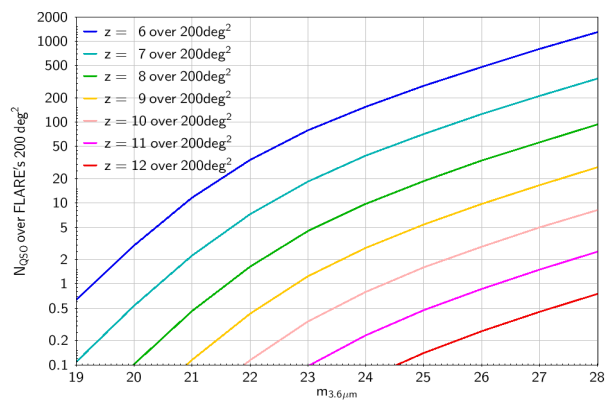

Figure 8: Extracted from Burgarella et al. (2016SPIE.9904E..2NB ). Basic predictions calculated by David Rosario using the Venemans et al. (2013) quasar luminosity function and the SDSS composite quasar spectrum for a 200 deg$^2$ survey. *SPACE* limiting magnitude will be m$_{AB}$ = 28.

Deep, broadband, wide-field photometric surveys such as the one presented in Section 2.2.1, provide a wealth of information on galaxies. However, this photometric information is not sufficient to gather a complete view of the formation and evolution of galaxies. Some physical information, such as the kinematic state of the galaxies and their chemical abundances, cannot readily be extracted from photometric data: spectroscopic observations are required. Furthermore, while continuum-bright galaxies can easily be detected using broad bands, part of the galaxy population is missed by these surveys: spectroscopic surveys without any prior selection do not demonstrate the same redshift distribution. Even spectroscopic follow-up of imaging surveys cannot provide a complete census of star-forming galaxies as some galaxies with extremely high equivalent width emission lines are undetected in broad-band imaging even with the deepest facilities (as shown by Bacon et al. 2015).

### 2.2.5. A wide-area integral field spectroscopic survey with *SPACE*: emission lines from high redshift star forming galaxies

The *SPACE* IFU will allow for the first time an unbiased deep survey for emission line sources over an area of ~1.5 deg$^2$ in total, many orders of magnitude larger than will be surveyed with the JWST NIRSpec IFU. Thanks to the NIR+MIR coverage of *SPACE* (1 - 28μm) and when they are bright enough, Hα and the other optical lines can be traced out to the first galaxies. The well-known rest-UV lines of Lyα, HeII1640, OIII]1663, CIII]1909 and CIV1550 can be observed for high-redshift sources (z > 5-7) and up to the largest distances where no galaxies have yet been found.

### 2.2.6. An Emission Line Survey at z > 7

Optical emission lines provide invaluable information on the chemical composition, ISM properties, nature/hardness of the ionizing source and on the instantaneous SFR, stellar age and related properties. But to collect a large number of objects needs wide and deep spectroscopic observations that are extremely difficult, even for JWST (de Barros et al. 2019, Wilkins et al. 2020). For sources with magnitudes brighter than m$_{AB}$=28 (detected in the continuum survey), the IFU survey will detect [OIII]5007 emitters with rest-frame Equivalent Widths (EW)~300Å (360Å) and higher at z ~ 7 (Fig. 9). Since the typical strength of [OIII] in star-forming galaxies at high-z is ~1000Å according to current estimates (Smit et al. 2014, Khostovan et al. 2016), *SPACE* should detect this line for all sources to the depth of the continuum survey (m$_{AB}$ = 28 for the *SPACE* imaging), corresponding to several hundred sources with [OIII] at z ~ 7 (Fig. 10). Furthermore, since fainter less massive galaxies generally have stronger emission lines, the IFU survey should discover pure emission lines sources undetected in the continuum surveys. This point is illustrated in Fig. 9 where a Lyα galaxy at z = 5.08 does not show any HST counterparts in the broad F606W and F814W bands. As many as 30% of entire Lyα emitter sample have no HST counterparts with I814W > 29.5.

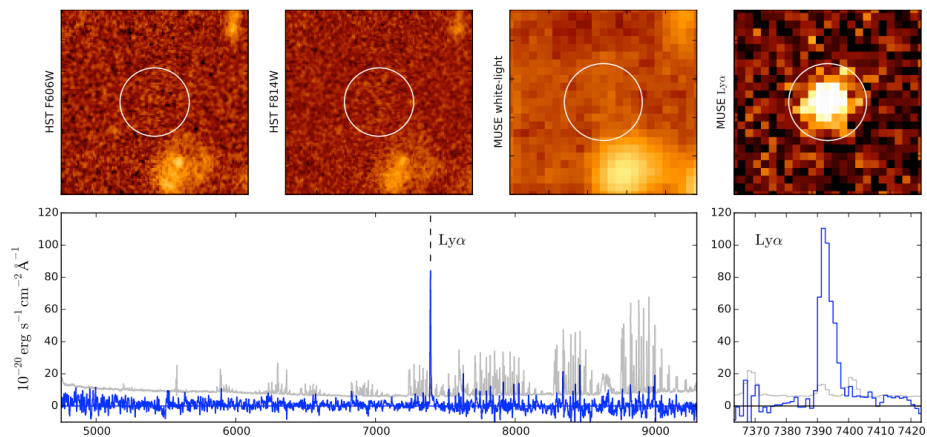

Figure 9: Integral-field spectroscopic observations with MUSE on the VLT (Bacon et al. 2015) in the Hubble Deep Field South showed that as many as 30% of entire Lyα emitter sample have no HST counterparts with I814 > 29.5. For instance, for this z = 5.08 object, HST broad bands are shown (top left), the MUSE reconstructed white-light and Lyα narrow band images (top right). The 1-arcsec radius circles show the emission line location. The full spectrum (in blue), smoothed with a 4 Å boxcar, and its 3σ error (in grey) are displayed below. A zoom of the unsmoothed spectrum, centered around the Lyα emission line, is shown (bottom right). Credit: R. Bacon / CRAL – Observatoire de Lyon.

The rest-UV emission lines of star-forming galaxies have lower equivalent widths than the optical lines. The strongest ones are Lyα (up to ~200 - 300Å for normal populations and higher for "exotic" cases, e.g. Pop III) and CIII]1909 with EW up to ~20Å (e.g. Stark et al. 2016). If EW(CIII]1909) ~ 10Å is typical, CIII] can be detected in fairly bright galaxies (m$_{AB}$ ~ 26-26.5 or brighter) at z ~ 7 to 9 with *SPACE*, and all the sources whose UV lines will be detected in the spectroscopic IFU survey (down to fluxes ~10$^{-18}$ erg/cm$^2$/s) will also be detected in the continuum survey. For these sources the line detection will provide spectroscopic redshifts and a key insight into the physical properties of galaxies including the hardness of the radiation field, chemical composition, ionizing photon production and others, as shown e.g. by recent pilot studies using ground-based near-IR spectroscopy (Stark et al. 2014, 2016) at z ~ 7 and recent models (e.g. Feltre, Charlot & Gutkin 2016, Nakajima et al. 2016).

### 2.2.7. The Evolution of the Star Formation Rate Density at z = 2-7

Selecting on Hα emission has been shown to be by far the most complete way to select star-forming galaxies. Due to its redder wavelength and typically high luminosity, selecting galaxies by their Hα emission can efficiently recover a large range of star-forming galaxies, from very blue to the most dusty/Sub-mm galaxies (e.g. Oteo et al. 2015). Furthermore, the Hα emission line may turn out to be even more important as a star formation indicator at higher redshift if star formation histories are significantly bursty.

*SPACE* will be the first mission to measure the Hα luminosity function (Fig. 10), and its evolution beyond z~2 and out to z~6-7. Most importantly, the spectroscopic/IFU capabilities mean that *SPACE* will be able to do this on a large volume (~1.5 deg$^2$), but also in a completely unbiased way without any pre-selection on galaxies being Lyman-break galaxies. *SPACE* will provide a unique insight into the overlap of purely Hα selected galaxies up to z~7 and those selected based on imaging and with the Lyman-break technique. How much star-formation are we missing? Are Hα and UV star formation rates consistent? Furthermore, these observations will also cover other emission lines, key to understanding the properties of high redshift galaxies. For these unique samples, the IFU will allow *SPACE* to also obtain spatially resolved information, including line ratios (for metallicity and dust extinction measures) and to obtain the dynamics (and hence masses) of high redshift galaxies.

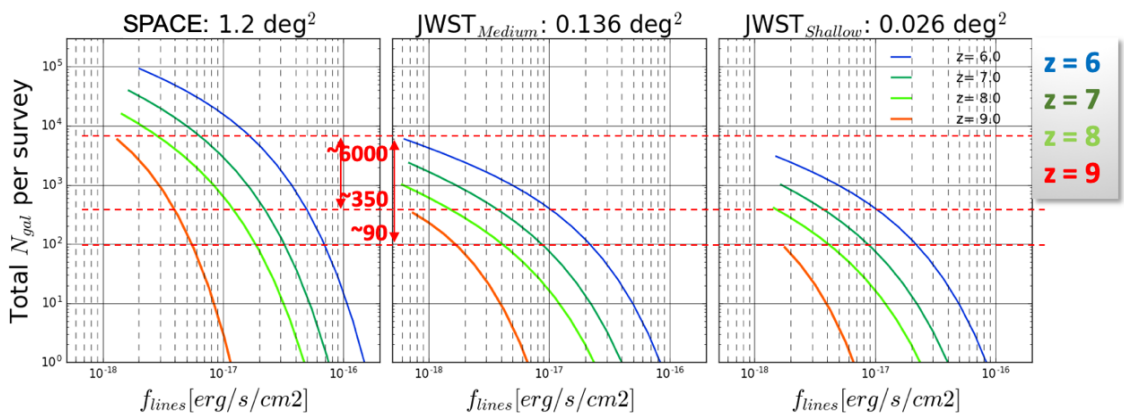

Figure 10: Using UV luminosity functions, we can estimate the number of galaxies that could be detectable by *SPACE* and by the JWST surveys (R > 1000). JWST Deep Spectroscopic survey is not plotted here because it is not in the same range. For the medium and deep spectroscopic surveys, we present the number of galaxies detected via their [OIII]5007 line (the same ratio is found for H-alpha and [OII]3727 but the redshift range is linked to the rest-frame wavelength of the line). At any redshift, *SPACE* will detect more than 10 times the number of objects that JWST could collect.

2.2.8. Searching for the First Generation of Stars with *SPACE* Spectroscopy

While several simulations and predictions have been made regarding the first generation of stars and black holes, including their hosts and redshift distribution (e.g. Yang et al. 2006; Schaerer 2003; Tornatore et al. 2007; Stacy et al. 2016), observational evidence has been lagging and failing to provide direct tests. This is essentially due to the lack of current instrumentation and telescopes to truly test them at high enough redshift. One needs to simultaneously probe high enough redshift, probe large enough volumes, have the sensitivity in the NIR, but also conduct surveys with an inclusive enough selection function such that extreme/interesting objects are not wrongly classed as stars or low redshift interlopers.

In practice, in order to truly make progress on what are expected to be rare, short-lived sources residing at very high redshifts, one clearly needs: i) near- to mid-IR coverage, ii) imaging and spectroscopic coverage and iii) capacity to cover the large enough volumes/areas that go significantly beyond the current state-of-the-art in an unbiased way. *SPACE* is the only mission that can combine all the necessary capabilities to make revolutionary discoveries on this end.

Prior to *SPACE*, JWST will have a very important role in further pushing this and providing the need to further develop tools to understand likely extreme stellar populations. However, during its limited, highly subscribed lifetime, it will mostly conduct spectroscopic follow-up of Lyman break selected galaxies. While some of those may turn out to be exciting sources, it is unlikely to be able to find the rare sources which only *SPACE* has the large enough volume to find. Furthermore, it is already clear that the search for pristine and/or extreme sources is not going to be simple at all. For example, one cannot simply detect high enough HeII equivalent width and be sure of the pristine nature of the source (e.g. Sobral et al. 2015; Hartwig et al. 2016). Regardless, it is also clear that the selection function may be key, and having an unbiased, blind IFU selection function in the key observable window (NIR and MIR) will provide the unprecedented survey capability to uncover tens to hundreds of sources similar to, e.g., CR7 (Sobral et al. 2015)..

On the other hand, several studies (e.g. Tornatore et al. 2007; Visbal et al. 2016) show that Pop III star formation, and direct collapse black holes (DCBH) form and remain well within the observable capabilities of *SPACE*. However, since we do not know their intrinsic spectra, nor their diversity (and that of their hosts), only a blind spectroscopic survey can truly find and study them (since some may have strange colours which may not fit into the standard broad-band Lyman break selection). *SPACE* will probe other lines apart from HeII, which is key to unveil the metallicity and physical conditions.

Within *SPACE*'s IFU survey, and with the current estimates for number densities of extreme sources based on either CR7 or e.g. Visbal et al. 2016 one would expect number densities around $10^{-5}$ to $10^{-6}$ Mpc$^{-3}$ around z~7 but rising at higher redshift. This means *SPACE* should be able to detect a truly statistical sample of ~100 of such sources without any pre-selection and to provide unique constraints on their nature directly by means of all the rest-frame UV lines detected.

3. Milky Way Studies: from the Global Structure to Dust Characterization

Stars form from interstellar cloud material. Massive stars form only in giant molecular clouds while solar-mass stars form in both small clouds and giant ones. Herschel has confirmed the prevalence of filaments inside interstellar clouds, along which prestellar cores and protostars could be preferentially seen (Fig. 11 from Stutz & Gould 2016 ; see also André 2017;). However, in detail we still do not understand a number of facts such as the role of the galactic environment on the formation and dynamics of molecular clouds ; the interplay between gravity, turbulence, and magnetic field in shaping the prestellar cores from cloud fragmentation ; the net effect of star-formation feedback due to supernovae, far-UV fluxes, and stellar winds from young (proto-)stars ; the relation between the masses of the clouds, of the cores and of the newborn stars, the formation of planets, the role of dust and ices in the clouds and in the water enrichment of the Earth, etc. Progressing on those open questions requires a multi-scale observational approach to characterize the interactions from the Galactic-structure down to the protostellar-disk scales, revealing dust properties and ice coverage in all clouds. With a large field of view, a reasonably high angular resolution, a near-to-far IR detector, *SPACE* will provide an opportunity for such multi-scale analysis on statistical grounds.

At large scales, the 3D distribution of the molecular clouds in the Galactic plane is challenging to infer. It is critical to understand the environmental effect on star formation, such as the impact of spiral arms or of the galactic bar on the molecular cloud formation and evolution. Moreover, distance estimates are essential to derive the cloud masses and identify the associated populations of young stellar objects (YSOs). Although the 3D structure of the ISM has already been explored, most methods rely on too simple assumptions and loosely-constrained ancillary data that limit the distance estimate, like the Galaxy rotation curve for kinematic distances or the dust temperature for inversion of dust emission maps.

The 3D structure of the ISM in the Galaxy will benefit from *SPACE* using a method free of the gas and dust biases mentioned above. The comparison of the star photometry with a model of stellar population synthesis (such as the so-called Besançon Galaxy model) permits to assign a distance and a reddening value to each star. The reddening is directly related to the column density which eventually yields a large-scale 3D map of the ISM. Previous attempts to build 3D extinction maps of the Galactic plane using 2MASS led to 10 arcmin resolution maps for the inner plane. Higher resolution can be achieved by including Gaia data, but only for the first few kiloparsecs, due to the strong extinction in the Galactic disk in the visible domain. Attempts to reach a higher spatial resolution with the deeper IR survey UKIDSS failed because in the near-IR spectral range the giant and dwarf stars cannot be separated upon their colour if they are too faint. *SPACE* with its 3 to 28 µm range and exquisite sensitivity will allow us to separate the two stellar populations and therefore build much better extinction maps. With deep infrared data from 3 to 28 µm, it will be possible to improve the spatial resolution to the sub-arcmin scale, reaching the arcsecond scale for clouds seen in absorption against the bright background in the mid-IR range, and probe lines of sight longer than 15 kpc. Such a map will be a goldmine for star formation investigations and more generally for Galactic structure studies.

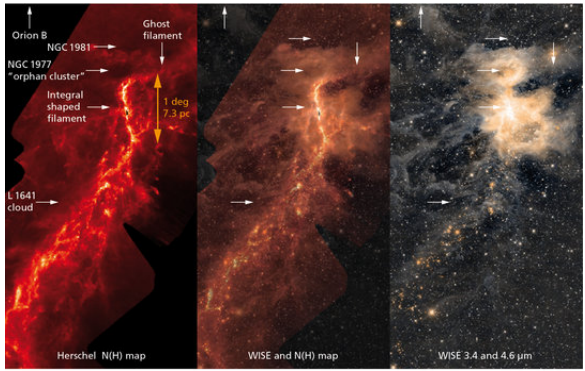

Figure 11: Extracted from Stutz & Gould (2016). Images of the Orion A star-forming cloud, showing the integral-shaped filament, the two-star clusters outside the filament, and the cloud L1641 to the South. Left: density map reconstructed from Herschel data, right: infrared image taken with the WISE space telescope (Lang 2014), center: combination of the two. Credit: A. M. Stutz / MPIA (now Univ. De Concepcion, Chile).

The study of star-forming regions (SFReg) spans many scales, from the molecular complex scale (a few tens of pc for the most massive ones), to the filament (a few pc), core (~0.1 pc) and disk (~0.001 pc) scales. It is a key to the understanding of star formation. Indeed, massive SFRegs play a pivotal role in the evolution of the Milky Way, and more generally of disk galaxies, since they are responsible for the formation of most stars in the Galaxy and are the only places where massive stars form. It has been understood early that the evolution of massive SFRegs is driven by the complex interplay between gravity, turbulence and magnetic fields, under the ambiguous influence of massive stars. The discovery of the ubiquity of filaments in the interstellar medium (ISM) has made it clear that these structures play a central role in how the gas flows from the diffuse large scales to the dense, star-forming cores. However, a major bottleneck is to disentangle the contributions of each physical process, because gravity and turbulence are multi-scale processes, and because they feed each other. Great progress was permitted in the last decade by numerical simulations, but a real breakthrough can only arise from a statistical comparison with observed SFRegs. However, observations remain hampered by the limited resolution and sensitivity of past and current facilities, which only enable one to fully characterize a handful of massive SFRegs. At 8μm, the 4m mirror of *SPACE* and its wide-field capabilities will enable us to map all the local massive SFRregs (Ménard, Scranton, Fukugita, & Richards 2010 @ <2 kpc) down to disk scales, and any SFReg of the Galaxy at resolutions < 0.1 pc.

In the mid-IR domain, surface brightness observations of massive SFRs will reveal the surface of clouds that are irradiated by local stars and by the interstellar radiation field (ISRF). Low-resolution (R~100-300) spectroscopic data in the 3-28 μm range will cover the emission of polycyclic aromatic hydrocarbons (PAHs) and very small grains (VSGs) which are mostly excited by far-UV photons. Past IR observatories have shown, on bright and nearby star-forming clouds, that the ratios of the aromatic spectral bands reflect the properties of the impinging radiation field. Along with the detection of the mid-IR $H_2$ excitation lines, stringent constraints can be put on the impact of nearby young and massive stars on their parent molecular cloud. *SPACE*'s wide-field and high-sensitivity performances will be crucial to generalize such analysis for many SFRs, since only the statistical comparison with star-formation indicators (e.g., the number and distribution of YSOs) will reveal the circumstances where SF feedback is positive or negative. Mid-IR $H_2$ excitation lines will also allow us to study shocked regions and outflows that extend on scales beyond JWST mapping capabilities. These are important retroaction activities of star formation which are crucial to understand since they regulate the star formation efficiency of galaxies.

Mid-IR point sources are a goldmine for massive SFReg studies. Colour-magnitude diagnostic tools, coupled with total extinction determined from Herschel far-IR imaging from the Hi-GAL survey, can pinpoint the YSO age distributions in clumps via detailed comparison with model isochrones. Even simple colour-colour diagrams can be used for broad evolutionary classification of YSOs, determining the SF history in each stellar cluster. The power of *SPACE* in this respect is its ability to do this for tens and tens of degrees of the Galactic Plane, enabling critical comparative studies of young cluster formation as a function of their location in spiral arms, inter-arm regions, proximity to star-formation triggering agents like HII bubbles and relatively evolved OB associations.

Analysing the number and colour distributions of point sources will also enable us to compute high-resolution extinction maps of massive SFRs. Used in combination with current (UKIDSS, VVV) and coming (WFIRST) near-IR deep surveys, the longer wavelengths of *SPACE* will make it possible to detect background stars through the high-column density parts of the molecular clouds. Absorption of the background light in the 8 to 20 μm range also allows to study the deeply extinct parts with a resolution only limited by *SPACE* capabilities.

Finally, point-sources will enable us to characterize the dense and populous stellar clusters that form in massive SFRs. The study of their mass spectrum as a function of the environmental conditions is key to understand which processes determine the mass and multiplicity of stellar systems, and the Galactic star-formation history. However, such studies are usually hampered by the incompleteness of the stellar census. The wide-field and high-sensitivity performances of *SPACE* in the mid-IR domain will guarantee the completeness of the stellar census and of the mass distribution while including the youngest and most deeply embedded YSOs that remain invisible at near-IR wavelengths (Fig. 12).

This will be possible down to the low-mass end. Low mass stars are the faintest but also most numerous objects in the Galaxy. Their large number makes them the most profuse planetary systems, and recent observational evidence tend to show that they preferentially harbour low-mass, Earth-like planets, rather than giant planets, including in their habitable zone. *SPACE* will be perfectly suited to detect these stars to complete their census down to the substellar regime and to the planetary mass objects as well. The initial mass function (IMF) is an important constraint of theoretical and simulation works, so that a fine characterization of the IMF is one way to address the long-standing debate on whether the fragmentation of interstellar clouds into prestellar cores is dominated by gravitational fragmentation or by turbulent fragmentation. Recent studies demonstrated that, due to the core external pressure resulting from turbulence, low-mass stars and brown dwarfs can be formed more efficiently in the case of turbulent fragmentation than with gravitational fragmentation. The high sensitivity combined with the mid-IR range is well adapted to construct a sample with good statistics of these faint and red objects. The large field of view will also allow *SPACE* to probe a whole range of environments for many clouds. It will enable us to see statistically if clouds with an active dynamics, (e.g. due to shock compression, or a high level of turbulence) are more prone to host low-mass object formation. Furthermore, the detection of the faint IR excesses around cool stars and brown dwarfs is a great challenge for upcoming instrumentation. Debris discs orders of magnitude fainter should be detectable around low-mass stars and brown dwarfs, up to a few hundred parsecs away. These observations will open new frontiers and will drastically increase our knowledge of how these enigmatic objects are formed and what kind of planets they can harbour.

Grains have been shown to grow. Far-IR emission, compared to extinction, can better be explained if bigger grains are considered and the 9.8µm silicate or the 3µm water ice line widths seem to widen on some lines of sight, which is again attributable to grain growth. The most direct proof of this growth has been the discovery of coreshine, the scattering of MIR light in the 3-5µm range from cloud cores, and beyond that, the scattering up to 8µm that partly compensates the diffuse light extinction seen in that wavelength range towards cloud cores. Scattering is thought to be a strong tool to constrain grain properties, and it has become crucial to understand the optical behavior of dust from 1 µm to 2000 µm. However, such models are still in their infancy and require deep and extensive data to be tested. Scattering is also seen as a tool to constrain the 3-dimensional (3D) structure of individual clouds when modelled from 1 to 10µm, each wavelength revealing a different layer of the clouds.

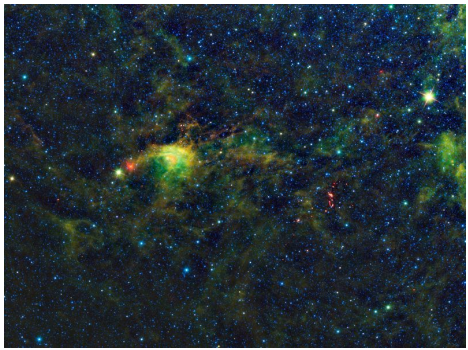

Figure 12: The Wide-Field Infrared Survey Explorer, or WISE, has uncovered a striking population of young stellar objects in a complex of dense, dark clouds in the southern constellation of Circinus. This mosaic from WISE covers an area on the sky so large that it could contain a grid of 11 by 7 full moons. The cloud itself is some 2,280 light-years away and spans more than 180 light-years across. *SPACE* will observe such fields at sub-arcsec resolutions. Image credit: NASA/JPL-Caltech / UCLA

Grains get covered with ices as soon as the clouds start to be shielded from UV ($A_V \geq$ 1-3 mag). Water ice is the first to appear but among the abundant ices, CO and $CO_2$ are two important ones (Fig. 13). Today, we do not know how they evolve inside the clouds: whether they totally evaporate around the newborn star and if they do, up to which distance. They modify the grain properties and therefore have an impact on the measurement of the cloud mass in parallel with the grain growth, which they also facilitate by increasing the sticking efficiency and resistance to shocks in turbulent collisions.

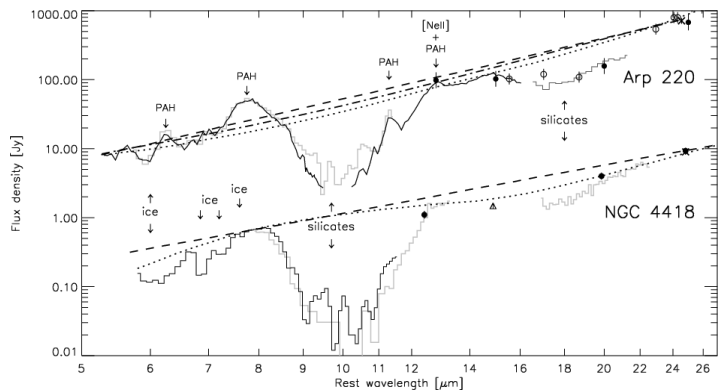

Figure 13: From Spoon et al. (2004). Comparison of the mid-IR spectra of Arp 220 (multiplied by 90) and NGC 4418. The positions of absorption and emission bands are indicated, as are several choices for the local continuum for each object. Credit: H. Spoon, Cornell Univ.

It is now clear that progress in the cloud studies today need to constrain the grain properties in parallel with the cloud structure and mass. To achieve such progress both absorption and scattering in the near and mid-IR need to be measured with the highest possible sensitivity on large surfaces. Combining such observations with far-IR emission from Herschel observations, we will assess both the structure of the clouds and the grain composition and size distribution in 3D. This will open up 3D radiative transfer studies of molecular emission, 3D chemistry and dynamical modelling. Like the Gaia project which, by measuring the stellar astrometry feeds many branches of astrophysics in dire need of stellar distances, *SPACE* mapping of the interstellar clouds will provide the basic data we strongly need to get the basic knowledge on the clouds from which all the physics of interstellar clouds, including their chemistry, will be completely revisited.

In the coming years, before the launch of *SPACE*, such studies will be carried on with the material we can collect presently but neither near-IR data from ground-based telescopes nor Spitzer archive data are sensitive enough to allow for the kind of studies we want to carry in the end. We will certainly make strong progress on the grain properties and start to produce 3D cloud models but with a limited spatial resolution by lack of sensitivity. Gaia observations can also be used to trace the 3D structure of the outer parts of the clouds but since it is working in the visible, it will not go beyond a few magnitudes of extinction. Ground-based telescopes can hardly observe the brightest objects in the mid-IR wavelengths and only in a few narrow windows. Spitzer is now retired, and ground-based telescopes will always be limited by the sky glow. No improvement is expected before the launch of *SPACE* except WFIRST which will provide the NIR coverage and the latter is therefore not included in this project. The JWST is of course very sensitive but its field of view is far too small to map thousands of square degrees.

## 4. Which Science Mission to Address the Science Questions?

The requirements derived from the science are summarized in Table 2 for a near-IR + mid-IR main instrument and in Table 3 for the (sub-)mm optional spectroscopic instrument.

With these requirements and with respect to the ESA Voyage 2050 programme, we have two options: one with a single standalone near-IR + mid-IR instrument in an M-sized mission. However, the science described in Sects. 1 and 2 would greatly benefit from a second option: a sub-mm instrument that would require an L-sized mission. This longer wavelength range allows to see inside the dustier regions and therefore contains a wealth of information on star formation. An all-sky survey to $3-5 \times 10^{-20}$ W/m$^2$ at $1\sigma$ in the range 300-1000μm would open a unique discovery space for a statistical study and to discover exotic objects. A similar instrument could perform a deep survey of about 1000 hours over our 200 sq. deg. with a sensitivity of about $1.3 \times 10^{-20}$ W/m$^2$ at 800μm.

### 4.1. What if we assume an M Mission?

*SPACE* could also be a standalone M-mission. The near-IR + mid-IR instrumental design would still be constrained by the wide-field requirement which is mandatory for our science case. We could also consider that *SPACE* could be co-funded by another space agency to reduce the total cost and finally meet the cost cap requirements.

One very important option that is also mentioned in another WP is that an instrument similar to *SPACE* (specifically MISC) could be an ESA contribution to a NASA project like the Origins Space Telescope (OST). Even though OST/MISC in the baseline design does not fully match our requirements, one of the OST up-scopes include most of what we need to carry out the science described in this WP.

### 4.2. What can we do with an L Science Mission?

*SPACE*'s scientific objectives would be better addressed if we assume an L mission with one imaging+spectrocopy instrument to cover the sky in the sub-mm and mm range. Such an instrument would allow to identify candidate very high redshift galaxies over the entire sky that we could follow-up with the near-IR + mid-IR instrument or to observe in parallel in near-IR + mid-IR while the telescope scans the sky.

Another option would be to build the mission in two phases. This would probably be more adapted to our science objectives. In the option described in Sect. 3.2, where we would share an L-mission with a project observing the sky in the (sub-)mm like, e.g., PRISM (Primordial Radiation Imaging and Spectroscopy Mission), an all-sky survey will be carried out first until $T = T_0$. Then at $T = T_1$, the mission will focus on the ecliptic poles to perform a deeper survey. The galactic disk might also be covered to shallower magnitudes. The alternative would be to observe everything in a parallel mode. *SPACE* would be one of the instruments.

| Size of the primary mirror | 4 – 6m |
|---|---|
| Wavelength range | 3 – 28 μm |
| Pixel scale imaging | 0.2 arcsec |
| Spectral resolution imaging | 3 - 5 |
| Instantaneous imaging field of view | 0.5 – 1 sq. deg. |
| Limiting flux mAB (1h, @ 5μm, SNR = 5) | 28 - 29 |
| Pixel scale spectroscopy | 0.4 arcsec |
| Instantaneous Spectroscopic field of view (MOS) | 0.5 – 1 sq. deg. |
| Instantaneous Spectroscopic field of view (IFS) | 1 sq. arcmin. |
| Spectral resolution (Low resolution) | 20 – 50 |
| Spectral resolution (Medium resolution) | 500 – 1000 |
| Limiting line flux (1h, @ 5μm SNR = 5) | $10^{-18} - 10^{-19}$ erg/cm2/s |

*Table 2: Summary of the requirements for the main near-IR + mid-IR instrument.*

| Size of the primary mirror | 4 – 6m |
|---|---|
| Wavelength range | 500 – 1000 μm |
| Pixel scale spectroscopy | Diff. limited |
| Spectral resolution (Medium resolution) | 300 |
| Limiting line flux (1h, @ 800μm, SNR = 5) | A few $10^{-20}$ W/m$^2$ |

*Table 3: Summary of the requirements for the optional (sub-)millimetric spectroscopic instrument.*

### 4.3. Technology Challenges

Several technological challenges are identified to build a facility for *SPACE*'s science. We provide details in the following of the section.

#### 4.3.1. Wide-field NIR+MIR detectors

In the NIR, up to about 5.3μm, we have now detectors that are quite efficient and space-qualified. For instance, JWST uses two different types of detectors: 4Mpixels H2RG detectors for the 0.6-5μm NIR (Teledyne Imaging Sensors) and 1Mpixel Si:As detectors for the 5-28 μm MIR (Raytheon Vision Systems). Both are built in California. So, we need to explore potential providers in Europe.

#### 4.3.2. Lightweight Large Mirror

For years the limit size for a lightweight space mirror, in the optical/IR range, was considered to be around 1.5m - 2.0m in diameter, mainly due to the difficulty to maintain the right shape of the mirror with the required precision at these wavelengths. A 2.0m+ Ø rigid enough mirror has an important mass that is a real drawback for a space mission while an ultra-lightweight mirror is then too floppy to maintain the optical quality of the telescope after launch and during operations. During the last years, developments have been pursued worldwide to break this limit, mainly in two directions:

- Large SiC mirror with excellent optical quality that allow large lightweight rigid mirrors,
- In-flight correction of thermo-elastic and gravity varying induced deformations of large lightweight mirrors using active loop with wave front sensors and deformable mirrors.

European industry as ADS/Mersen-Boostec has a real expertise with the manufacturing of the 3.5m mirror of the Herschel telescope (Sein et al. 2003), but excellent results have been published by Zhang et al. (2019, Fig. 14) demonstrating the manufacturing of a 4m SiC lightweight mirror with a surface shape better than λ/30 in optical wavelength (λ=632.8nm). For future large missions, the technology for manufacturing large primary mirrors is ready and will be soon mature.

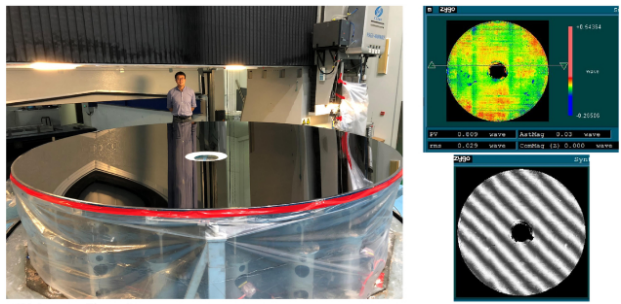

Figure 14: Light-weight mirror by Ge Zhang et al. (2019) that shows that the technology makes fast progress that suggests that our present assumptions are realistic objectives for the Voyage 2050 programme.

Credit: G. Zhang / Changchun Institute of Optics, Fine Mechanics and Physics (CIOMP)

### 4.3.3. Deformable Mirror for Active Optics

Only in Europe, many developments on deformable mirrors for space active optics have been pursued during the past 10 years (Fig. 15. In Europe, ESA has recently funded a number of programmes to further develop DM technologies specifically for space applications. Münster University developed a small-format Unimorph post-focal DM. As part of two independent active correction chain projects (i.e. WFSensing, DMs, Algorithms), two large monolithic deformable mirrors were also developed. The first, AOCC, was developed by TNO and the second, STOIC, by Fraunhofer IOF in collaboration with NUI Galway. Other European and national agencies have funded DM developments over the years; either to push forward new concepts or increase the TRL of already developed and used technologies. In France, Thales-Alenia-Space space validated a new active DM, based on a concept defined with LAM. This MADRAS DM, developed at TAS under CNES contract, sustains launch environment without locking device. The device has been fully integrated, and performance successfully verified. It is space qualified before assembly on the full-scale active optic instrument demonstrator.

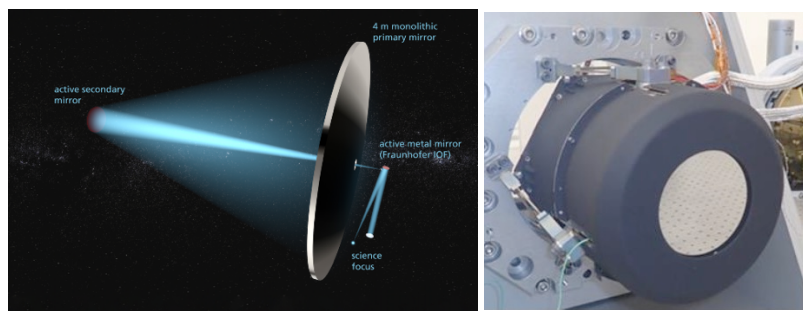

Figure 15: Left: STOIC concept with 4m monolithic mirror and active deformable mirrors © Fraunhofer IOF. See: https://exoplanets.nasa.gov/exep/technology/TDEM-awards/). Right: MADRAS DM © Thales-Alenia-Space & LAM.

Finally, a large number of traditional DM manufacturers are also present in Europe (e.g. Cilas, ADS-Microgate, Alpao, Imagine Optic). In the USA, NASA funded a number of DM technology developments for space applications, mainly through the Technology Development for Exoplanet Missions (TDEM, in particular companies such as, Boston Micromachines Inc (BMM) and Xinetic).

### 4.3.4. Wide-field Integral-Field Spectrograph

Integral field spectroscopy has proved to be a very powerful tool in recent years for studies of the distant universe using ground-based facilities. The combination of spatial and spectral information allows key diagnostics of galaxy formation and assembly to be probed, such as the kinematics of disk formation and the growth of galaxy bulges and black holes. Narrow-field integral field capabilities are planned for JWST, but the wide-field IR integral field capability proposed for *SPACE* is unique. It will open up a completely new discovery space which is inaccessible from the ground. In addition to the targeted spectroscopic surveys outlined in previous chapters, integral field spectroscopy has a unique potential for serendipitous discoveries by data-mining the resultant 3D data cubes for new emission-line sources invisible at other wavelengths. The technology for producing image slicers for integral field spectroscopy using diamond-machining techniques is now mature, and scalable from smaller field prototypes developed for NIRSpec and MIRI on JWST (Laurent et al., 2004). For instance, a method for cost efficient and high performant manufacturing of spherical image slicers was developed and patented in collaboration with the Winlight company. The technology was applied to VLT-MUSE instrument and is proposed for E-ELT and space instruments (Fig. 16).

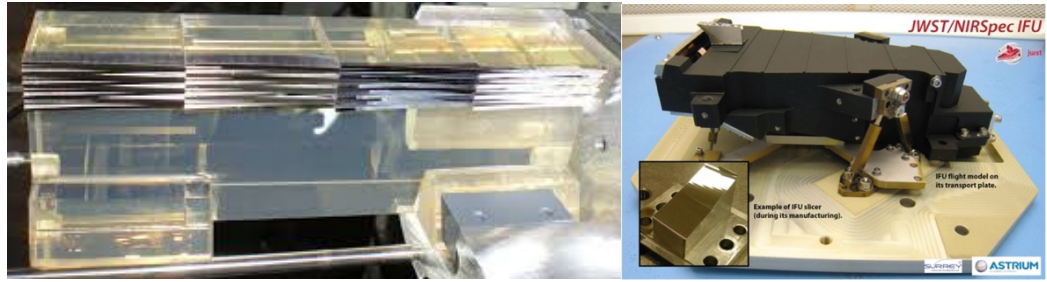

Fig 16: Left: The Image Dissector Array (IDA) manufacturing constitutes a WINLIGHT Optics/CNRS patent. Innovative methods, developed conjointly by LAM (Laboratoire d'Astrophysique de Marseille, France) and WinLight Optics (Marseille, France), allow reaching high performances (accurate roughness, sharp edges, surface form, etc.) with standard glass manufactured components while saving costs and time by an order of magnitude compared with classical techniques. This IDA is constituted by 48 slices. Right: Main picture: The NIRSpec IFU flight model. The IFU is roughly the size of a shoe box and weighs less than 1 kg. Insert: A diamond-machined image slicer, a key element of the IFU, pictured during manufacture. Credit: R. Sharples / University of Durham, United Kingdom.

### 4.3.5. Micro-Mirror Arrays

The scientific return from future Astrophysical space missions could be optimized using MOEMS (Micro-Opto-Electro-Mechanical systems) devices like large micromirror arrays (MMA). Multi-object spectrographs (MOS) are powerful tools for space and ground-based telescopes for the study of the formation and evolution of galaxies. This technique requires a programmable slit mask for astronomical object selection; 2D MMAs are suited for this task. MOEMS has been used to build JWST NIRSpec. In Europe, several options exist as the one from Laboratoire d'Astrophysique de Marseille (LAM, France) and the Centre Suisse d'Electronique et de Microtechnologies (CSEM, Switzerland) engaged in a European development of MMAs, called MIRA, exhibiting remarkable performances in terms of surface quality as well as ability to work at cryogenic temperatures. MMA with 100 x 200 $\mu m^2$ single-crystal silicon micromirrors were successfully designed, fabricated and tested down to 162 K (Fig. 17). They are designed to work at 30K and there are no blocking points that would prevent them to work at 5K. In order to fill large focal planes (mosaicking of several chips), we are currently developing large micromirror arrays to be integrated with their electronics.

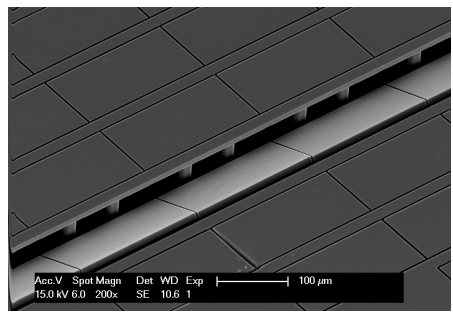

Fig. 17: Micro-mirror array with high fill factor in the long direction providing long slits. Each mirror measures 200 x 100 $\mu m^2$. Our project will take advantage of the already available building blocks to design, realize and package customized micro-mirror arrays perfectly suited for our instruments. 2D arrays are built on wafer with Through Wafer Vias in order to allow routing of the device on wafer backside, foreseeing integration with dedicated ASICs. Like for CCDs, mosaicking will permit wide fields of view. Credit: F. Zamkotsian / LAM & CSEM.

### 5. Conclusion and Roadmap

We present in this paper a scientific project to study the early universe, from the first stars to the end of the reionization. To meet the science objectives, we define a Space Project for Astrophysical and Cosmological Exploration (SPACE) as part on the ESA long term planning Voyage 2050 programme. SPACE will chart the formation of the heavy elements, measure the evolution of the galaxy luminosity function, trace the build-up of stellar mass in galaxies over cosmic time, and find the first super-massive black holes (SMBHs) to form.

SPACE will provide us with a much larger sample of objects in the early universe than any current or planned mission of very high redshift galaxies at z > 10. These objects will be bright enough for a detailed follow-up spectroscopy.

To our knowledge, no single planned project can directly address the scientific objectives detailed in this paper for several reasons:

- To carry out the rise of metals science case, we must observe in the range $3 - 28\mu m$ to follow the bright rest-frame optical lines and/or the PAH3.3$\mu m$ feature. No other tracers can measure the metallicity of galaxies at z $\gtrsim$ 5 efficiently and reliably.
- For all the science cases described in this WP, no planned instrumental project is optimal. We need a field of view much wider that the ones provided by e.g., JWST, SPICA/SMI and even the present OST/MISC. On the other hand, AKARI and WISE all-sky surveys are not deep enough, and their angular resolutions are too low.
- Ground-based telescopes in the optical+NIR like the ELT, in the (sub-)mm like ALMA, NOEMA, LMT are not competitive because of the wavelength range (ELT), the field of view (ALMA, NOEMA) and because spectroscopy in not integrated/developed (e.g. NIKA2, LMT). *SPACE* will be complementary to the SKA that will provide us with the observations of anisotropies in the brightness temperature of the 21 cm line of neutral hydrogen from the period before reionization and would shed light on the dawn of the first stars and galaxies (Santos et al.2011, A&A 527, 93).